\documentclass[preprint,12pt]{elsarticle}
\usepackage{graphicx}
\usepackage[figurename=Fig.,bf]{caption}
\usepackage{epsfig}
\usepackage{subfig}
\usepackage{amsmath}
\usepackage{physics}
\usepackage{natbib}

\begin{document}
	\begin{frontmatter}	
		\address[famu]{Florida A\&M University, Department of Physics, Tallahassee FL, 32307}

		\title{Electron gases in toroidal shells: mode coupling and state functions}		
		\author [famu]{M. Encinosa\corref{cor1}}
		\ead{mario.encinosa@famu.edu}
		\author [famu]{J. {Williamson}}
		\ead{johnny1.williamson@famu.edu}
		
		\cortext[cor1]{Corresponding Author}
		
		\begin{abstract}
			Eigenvalues and wave functions describing free electron gases in toroidal shells are determined using a basis set expansion natural to the system geometry. Couplings between azimuthal and poloidal modes are found to be appreciable at lower values of ${m_\phi}$, modifying single particle density distributions within the shells. Thermodynamic state functions are calculated and contrasted with those of a solid toroidal volume.  
		\end{abstract}

		
	\end{frontmatter}

	\section{Introduction}
	Nanoscale shape engineering has emerged as a vibrant subject of investigation within several areas of physics \cite{{juhan1},{aydin1}, {aydin2}}. As fabrication methods become increasingly sophisticated, the geometry of nano-materials is no longer synonymous with only their overall shapes, but is also now inclusive of periodic corrugations \cite{{cheng1},{cheng2}}, ripples \cite{{ripples1},{qdhe}} and twist distortions \cite{chirality1}.  
	
	An intriguing question that arises for nano-systems with novel configurations concerns how their properties are affected by curvature and confinement mechanisms. Toroidal structures have been, and likely will continue to be, subjects of study due to their potential applications as traps \cite{{gold},{trap2}} for cold molecules, biosensors \cite{biosensor1, biosensor2}, and qubits \cite{silva1}.  Additionally, torodial structures provide laboratories for exploring local curvature\cite{{silva2},{scriptaenc}, {encmag}} and flux quantization effects. In principle a magnetic field through a conducting ring, torus or any genus one object can induce persistent currents (Gauss-Laguerre beams may serve as means of generating poloidal modes giving rise to toroidal moments in extended objects \cite{tormoment1}).  
	
	In \cite{bhks}, the Schr{\"o}dinger equation (SE) for  spinless charge carriers in a toroidal volume (TV) was derived and solved using finite element methods on a disk within the volume; azimuthal symmetry then allowed for the solution to be known everywhere. In \cite{WE}  solutions for the same system were found by reducing the SE to single variable form with effective potentials accounting for the coupling between radial, poloidal and azimuthual degrees of freedom. In this work, the emphasis  is shifted towards understanding the properties of an electron gas confined within a toroidal shell (TS), anticipating circumstances in which analytic expressions for wave functions may be of importance. 
	
	This paper is organized as follows: Section 2 presents a brief review of the derivation of the TV SE and describes the basis set used in this work.  Section 3 presents spectra and wave functions, and illustrates the rapidity with which toroidal degrees of freedom begin to display substantial coupling. Eigenvalues are employed to calculate the heat capacity, free energy and degeneracy pressure  within a shell and are compared to those calculated for a TV. Section 4 presents conclusions and some remarks regarding extensions of this work. 
	
	\section{The Hamiltonian and basis set}
	The SE for a shell is the same as for a volume, however the solutions are modified due to the different boundary conditions. To keep this work self-contained, the derivation of the SE for a solid torus is given below. 
	
	A solid torus with major radius $R$ minor radius $a$, poloidal angle $\theta$, and azimuthal angle $\phi$ is defined by the set of points ${\textbf r}(\rho,\theta,\phi)$ which satisfy
	
	\begin{equation} \label{eq:monge}
		{\textbf r}(\rho,\theta,\phi)=W(\rho,\theta)\,\hat{\boldsymbol{\rho}}+\rho\, \sin\theta\,\hat{\textrm{\textbf{k}}},
	\end{equation}

	\noindent where
	
	\begin{equation} \label{eq:W}
		W(\rho,\theta)=R+\rho\,\cos\theta
	\end{equation}
	
	\noindent
	The metric and Laplacian can be derived from  Eq.\,(\ref{eq:monge}) by standard methods \cite{encinosaT2}. However, results will be shown in section 3 that include magnetic field effects, so it is preferable to first write the gradient. In terms of unit vectors $\hat{{n}}$ (normal to a toroidal surface at fixed $\rho$), $\hat{\theta}$ (tangent to the surface defined by constant $\rho$) and $\hat{\phi}$ in the azimuthal direction,  the gradient is
	
		\begin{equation} \label{eq:grad}
		\boldsymbol{\nabla} =\hat{\boldsymbol{\rm n}} \frac {\partial} {\partial \rho}+
		\hat{\boldsymbol{\theta}} \frac{1}{\rho} \frac{\partial}{\partial \theta}+
		\hat{\boldsymbol{\phi}}  \frac{1}{W(\rho,\theta)} \frac{\partial}{ \partial \phi}
	\end{equation}

	\noindent which will be used in the time-independent Schr\"{o}dinger equation (in the Coulomb gauge) 
	\begin{equation} \label{eq:schrod}
		\bigg( \nabla ^2 + 2 i \frac{e}{\hbar}~\textbf{A} \cdot \boldsymbol{\nabla} - \frac{e^2}{\hbar^2} \textbf{A}^2 +  \frac{2  m_e E}{\hbar^2}  \bigg) \Psi = 0,
	\end{equation}
	where $e$ and $m_e$ are the electron charge and mass. It proves convenient to define the parameters
	\begin{equation*} \label{eq:defs}
		\begin{split}
			u =&\, \frac{\rho}{R}\\
			\alpha =&\, \frac{a}{R}\\
			\gamma_0 =&\, B_0 \pi R^2 \\
			\gamma_{\mbox{\tiny N}} =&\, \frac{\pi \hbar}{e}\\
			\epsilon =&\, \frac{2 m_e E R^2}{\hbar^2}\\
			\tau_0 =&\, \frac{\gamma_0}{\gamma_{\mbox{\tiny N}}}\\
			W(u,\theta) =&\, 1+u \cos\theta,
		\end{split}
	\end{equation*}
	\noindent
	after which, expansion of  Eq.\,(\ref{eq:schrod}) takes the dimensionless form
	
	\begin{equation} \label{eq:dimensionless}
		\begin{split}
			\bigg[\frac{\partial^2}{\partial u^2} + & \frac{1}{u}\frac{\partial}{\partial u}  + \frac {1} {u^2} \frac{\partial^2}{\partial \theta^2}+  \frac {\cos\theta} {W(u,\theta)} \frac{\partial}{\partial u} + \frac {1}{W^2 (u,\theta)} \frac{\partial^2}{\partial \phi^2}  \\ & - \frac {\sin\theta} {u W(u,\theta)} \frac {\partial} {\partial \theta} -  \frac {\tau_0} {i} \frac {\partial}{\partial \phi} - \frac {1} {4} \tau^2_0 W^2(u,\theta) + \epsilon \bigg]\psi = 0
		\end{split}
	\end{equation}
	
	\noindent
	
	The three leading terms in Eq.\,(\ref{eq:dimensionless}) comprise part of Bessel's equation. The $\theta \rightarrow -\theta$ invariance of  Eq.\,(\ref{eq:dimensionless}) allow for its solutions to be separated into even  $\rm cos[n\theta]$ and odd $\rm sin[n\theta]$ parities, providing  a natural choice for a basis expansion in terms of Bessel functions of the first kind,
	
	\begin{equation} \label{eq:basis}
		\chi{^{\pm}_{\bar{n} \bar{\nu} m_\phi}} (u,\theta)=\sum_{n\nu}\frac {1}{b_{n\nu} t_n} C{^{n\nu}_{\bar{n} \bar{\nu} m_\phi}}J_n \biggl (\frac{x_{n\nu}u} {\alpha} \biggr)\begin{pmatrix}\cos n\theta \\ \sin n\theta \end{pmatrix}
	\end{equation}
	
	\noindent
	with $b_{n\nu}$ and $t_n$ Bessel function and trigonometric normalizations.
	
The first order derivative in $\theta$ term can be eliminated with an integrating factor. The basis set elements then have the form
	\begin{equation} \label{eq:aux}
		\psi^{\pm}_{\bar{n} \bar{\nu} m_\phi}(u,\theta,\phi)=\frac {1} {\sqrt{W(u,\theta)}} \chi^{\pm}_{\bar{n} \bar{\nu} m_\phi}(u,\theta)\exp[i m_\phi \phi]
	\end{equation}
	Only one term in Eq.\,(\ref{eq:dimensionless}) is not separable into a product form in $u$ and $\theta$. In \cite{WE} the $\theta$-dependence of this  term was integrated out by contour methods to yield effective potentials in the radial variable  defined here as $\Lambda_{\bar{n} n}^{\pm}(u)$. For a given ${m_\phi}$ the matrix elements reduce to 
	\begin{equation} \label{eq:elements}
		\begin{split}
			H_{\bar{n}\bar{\nu}, n \nu} =&  - \bigg [ \biggl (\frac {x_{n\nu}} {\alpha}\biggr )^2 + \tau_0 m_\phi + \frac {\tau_0^2} {4} \bigg ] \delta_{\bar{n} n} 
			\delta_{\bar{\nu} \nu} \\ &- \bra{\bar{n}\bar{\nu}} (\frac{1} {4}-m_\phi^2) \Lambda_{\bar{n} n}^{\pm}(u)+  \frac {1} {2}\tau^2_0 u  f^{\pm}_1(n ,\bar{n}) + \frac {1} {8}\tau^2_0 u^2  f^{\pm}_2(n, \bar{n})\ket{{n \nu}}
		\end{split}
	\end{equation}
	where
	\begin{equation} \label{eq:fnnbar}
		f^{\pm}_k(\bar{n} ,n) \equiv \int_{0}^{2\pi} d\theta  \begin{pmatrix} \cos \bar{n}\theta \cos k \theta  \cos n\theta \\ \sin \bar{n} \theta \sin k \theta  \sin n\theta \end{pmatrix}
	\end{equation}
	\noindent reduces to a sum over Kronecker delta functions in $\bar{n}, k$, and $n$. Explicit forms for $\Lambda_{\bar{n} n}^{\pm}(u)$ are presented in the appendix.

	The outer wall of the torus occurs at $u = \alpha$.  To create a shell, define $0 < s  \leq 1$ as the parameter that sets the position of its interior wall. Excluding the origin $u = 0$ requires 
	
	\begin{equation} \label{eq:shell}
		J_n \bigg( \frac {x_{n\nu}u} {\alpha}\bigg)\rightarrow A_{n \nu} J_n \bigg( \frac {\eta_{n\nu}u} {\alpha}\bigg)+B_{n \nu} Y_n \bigg( \frac {\eta_{n\nu}u} {\alpha}\bigg) \equiv T_n \bigg( \frac {\eta_{n\nu}u} {\alpha}\bigg) 
	\end{equation}
	\noindent
	with $Y_{n}$ a Bessel function of the second kind.  It is then a standard exercise to obtain the  $\eta_{n\nu}$ appearing in Eq.\,(\ref{eq:shell}) from (suppressing indices for clarity) 
	\begin{equation}
		A J_n (s\alpha) + B Y_n (s\alpha) = 0
	\end{equation}
	\begin{equation}
		A J_n (\alpha) + B Y_n (\alpha) = 0
	\end{equation}
	\noindent
	The ratio  $B/A$ of irregular to regular solutions can be obtained for a given $\alpha$ satisfying Eqs. (11) and (12).
	
	In this work, 100 basis functions were employed to find eigenvalues and wave functions for both positive and negative parities. There are no terms  Eq.\,(\ref{eq:dimensionless}) that mix parity states. The positive state basis set comprises functions with $n=(0,1,2,3,4)$  coupled to the first twenty associated zeros $\nu = (1,2,...,20)$  found by solving Eqs. (11) and (12). The negative parity states have $n=(1,2,3,4,5 )$. In accordance with and to facilitate comparison with previous work \cite{{bhks},{WE}}, a GaAs effective mass $m = .067 m_e$ is assumed.

	\section{Results}
	In Table \ref{tab1},  representative zeros of $T_n$ for aspect ratio $\alpha = 0.5$ and $s = 0.25$, $s = 0.5$,  $s = 0.75$ are shown next to multiples of $\pi$.  Inclusion of the irregular Bessel function leads to the $\eta_{n\nu}$ being reasonably approximated with a hard wall form $\sim n \pi / (1-s)$ even for an thicker shell ($s = 0.25$). This observation may prove useful for developing approximations that circumvent finding roots of the $T_n$. Table \ref{tab2} illustrates  $B_{n\nu}$  coefficients for the Table \ref{tab1} basis states normalized to $A_{n\nu} = 1$. Patterns that emerge among the $B_{n\nu}$ cluster into the same $n$-values of the $\eta_{n\nu}$  and are only useful as a rough consistency check. 
	
	Tables \ref{tab3} and \ref{tab4} contain examples of eigenstates generated from solving Eq.\,(\ref{eq:dimensionless}) in terms of the basis at a given $m_\phi$ for the full ($s = 0$)  torus. Two representative states were selected to illustrate the general behavior observed within the substantially larger set of results.  It was expected the lowest energy eigenstates would be substantially altered as  $m_\phi$ increased, but as evidenced in Table 4, this was generally not the case. Tables \ref{tab5} and \ref{tab6} are results for an $s = 0.75$ shell. The interplay of confinement and mode coupling within a region corresponding to a thin film of thickness 12.5 nm is substantial, and appears to be a consequence of the $\Lambda_{\bar{n} n}(u)$ potential assuming its maximum values (of order unity) near $u \sim \alpha$. In contrast, for the full torus, the $J_n$ basis states overlap near $\rho \sim 0$ where $\Lambda_{\bar{n} n}(u)$ is smaller and are oscillatory over $0 \leq s \leq \alpha$. 
	
\begin{figure}
		\centering
		\includegraphics[width=0.6\columnwidth]{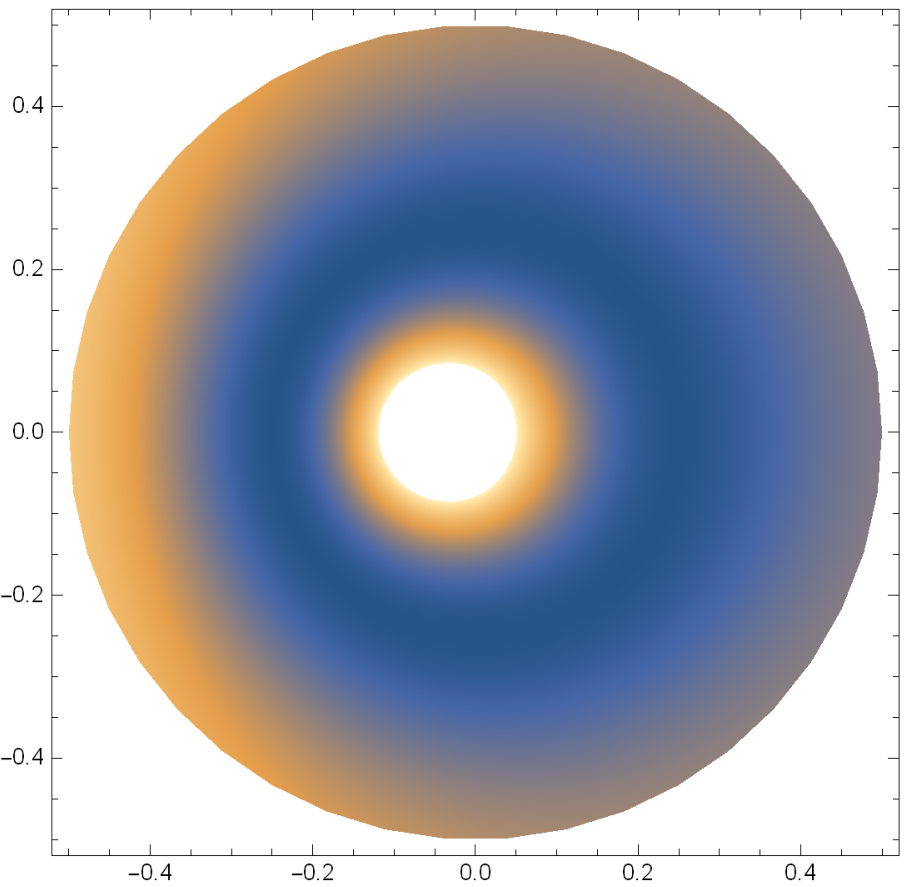}
		\caption{$\vert \psi{^{+}_{0 1 0} (u,\theta)} \vert^2$ for the full torus with $\alpha = 0.5$.}
		\label{fig:Dens1}
	\end{figure}

\begin{figure}
		\centering
		\includegraphics[width=0.6\columnwidth]{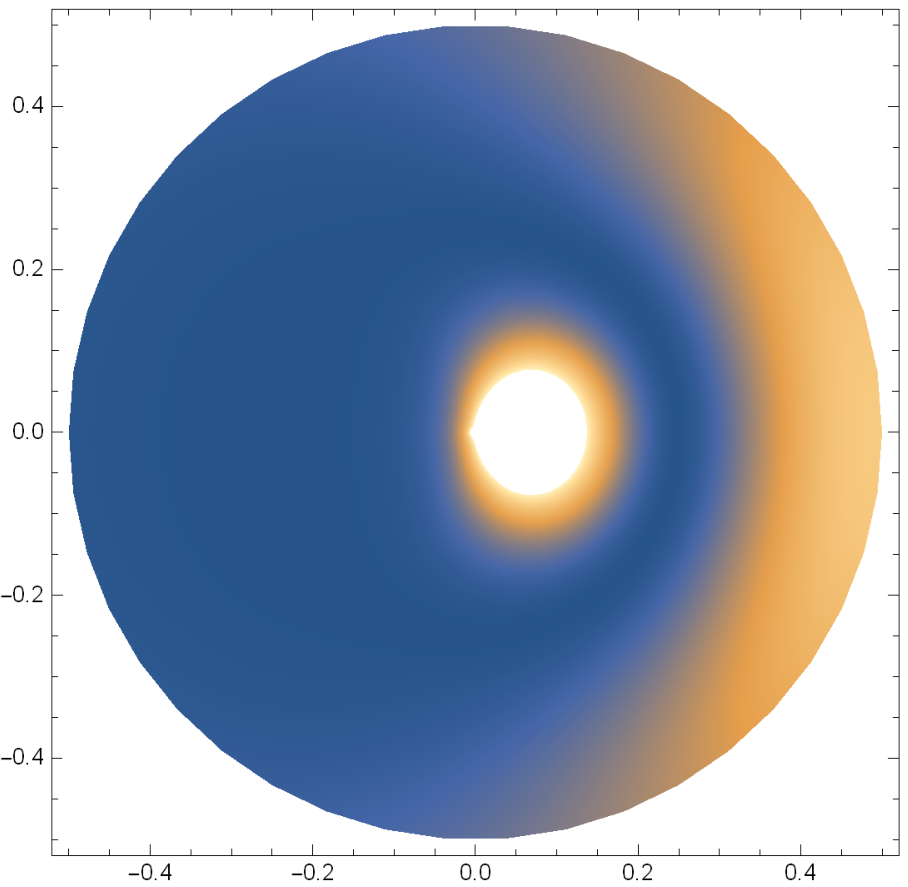}
		\caption{$\vert \psi{^{+}_{0 1 (10)} (u,\theta)} \vert^2$ for the full torus with $\alpha = 0.5$.}
		\label{fig:Dens2}
	\end{figure}

\begin{figure}
		\centering
		\includegraphics[width=0.6\columnwidth]{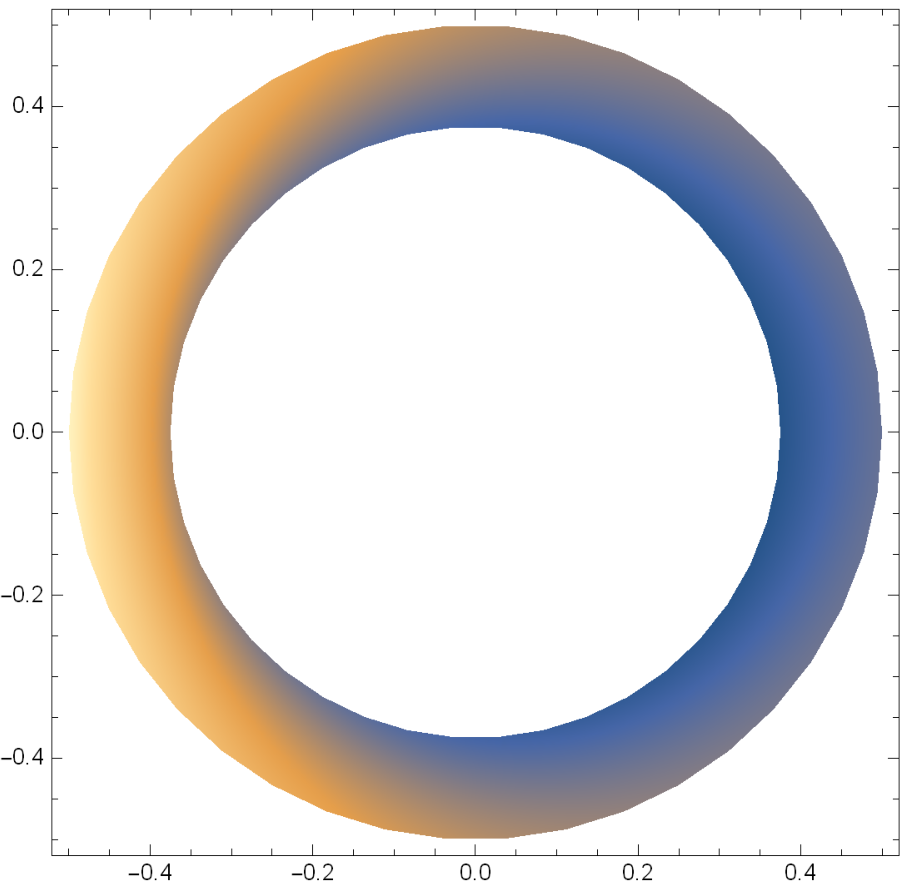}
		\caption{$\vert \psi{^{+}_{0 1 0} (u,\theta)} \vert^2$ for shell $s = 0.75$, $\alpha = 0.5$.}
		\label{fig:Dens3}
	\end{figure}

\begin{figure}
		\centering
		\includegraphics[width=0.6\columnwidth]{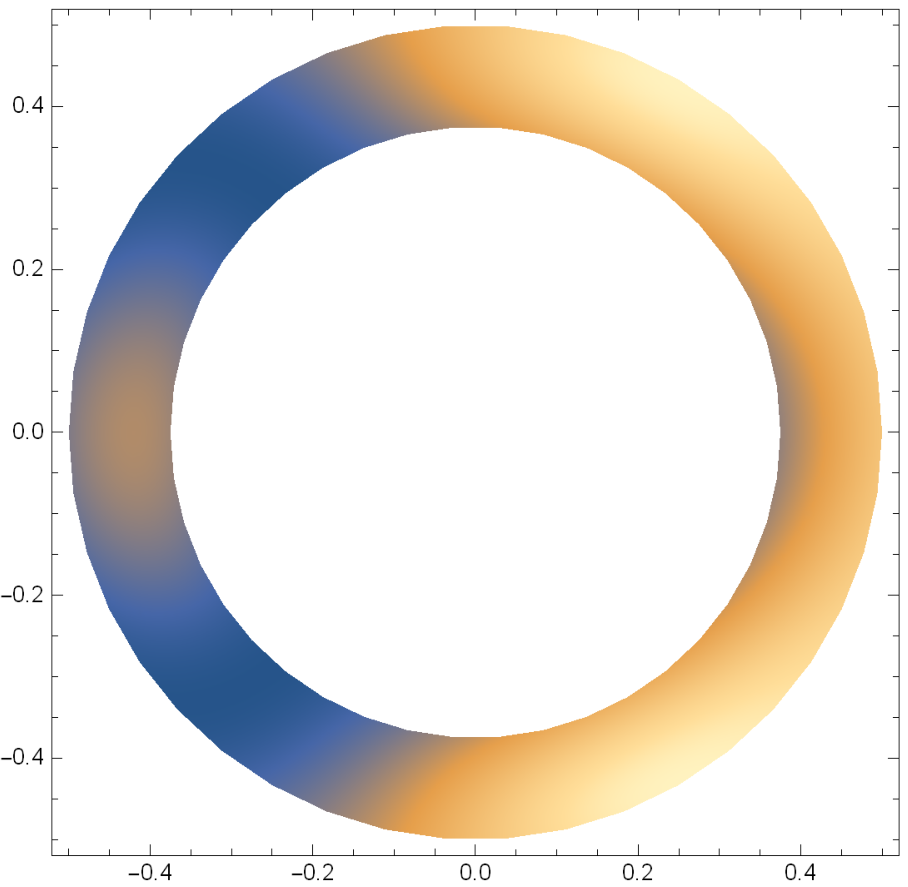}
		\caption{$\vert \psi{^{+}_{0 1 (10)} (u,\theta)} \vert^2$ for shell $s = 0.75$, $\alpha = 0.5$.}
		\label{fig:Dens4}
	\end{figure}

\begin{figure}
		\centering
		\includegraphics[width=0.6\columnwidth]{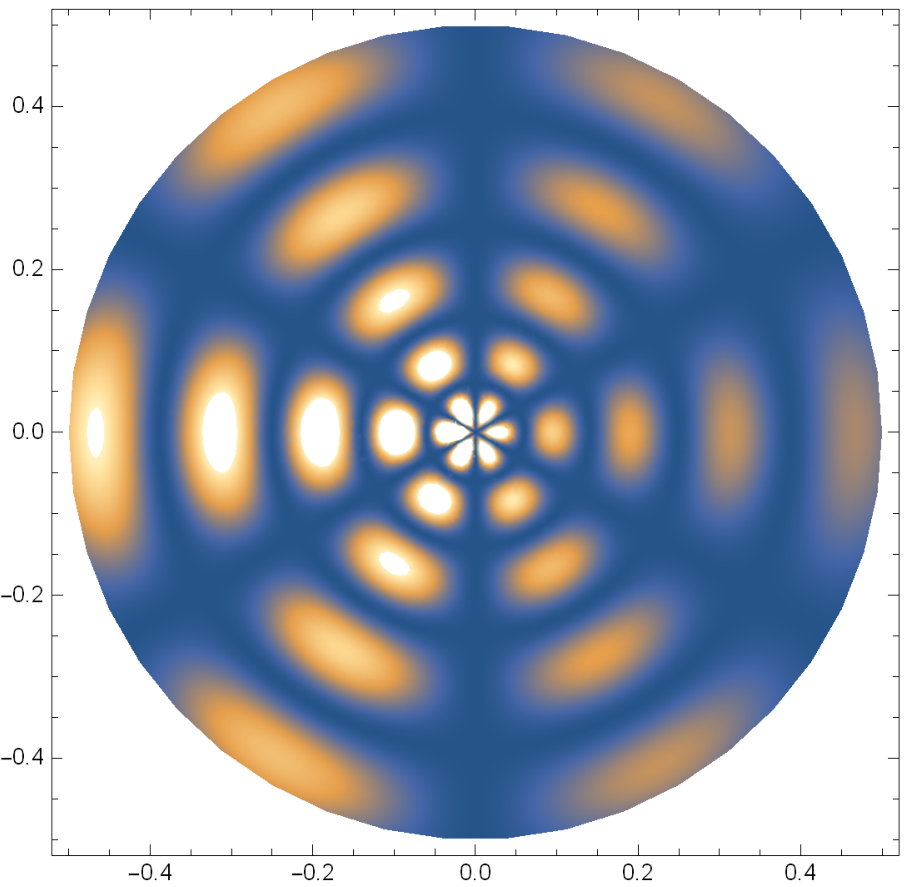}
		\caption{$\vert \psi{^{+}_{3 3 0} (u,\theta)} \vert^2$ for the full torus, $\alpha = 0.5$.}
		\label{fig:Dens5}
	\end{figure}

\begin{figure}
		\centering
		\includegraphics[width=0.6\columnwidth]{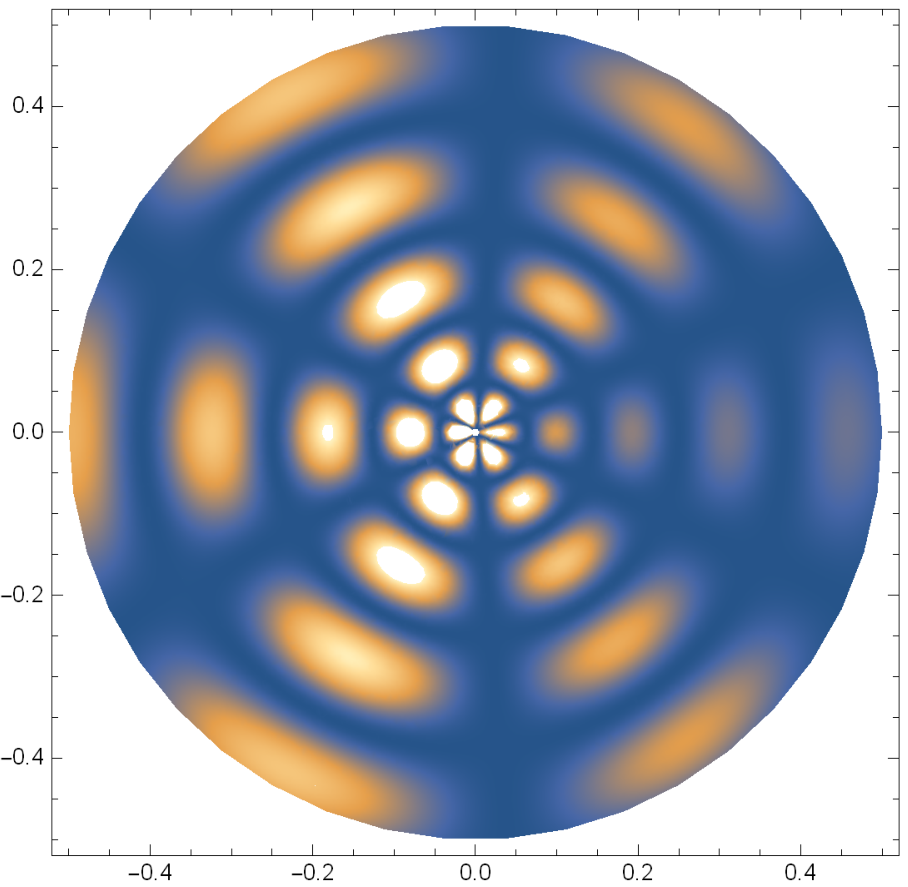}
		\caption{$\vert \psi{^{+}_{3 3 (10)} (u,\theta)} \vert^2$ for the full torus, $\alpha = 0.5$.}
		\label{fig:Dens6}
	\end{figure}

\begin{figure}
		\centering
		\includegraphics[width=0.6\columnwidth]{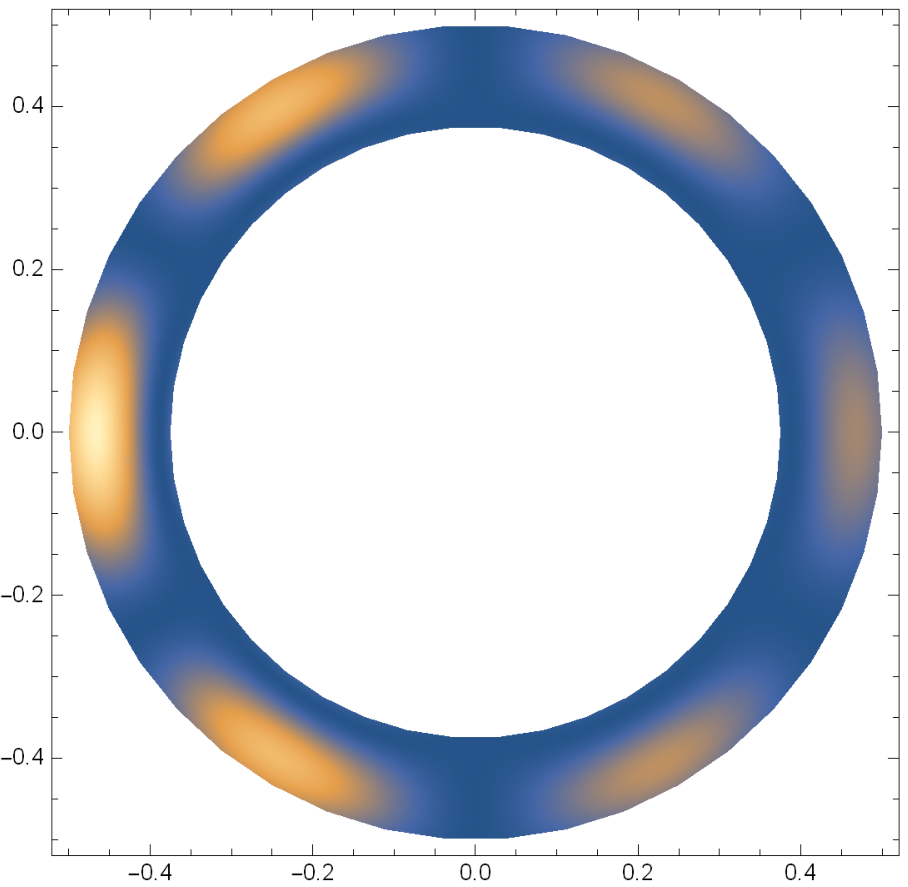}
		\caption{$\vert \psi{^{+}_{3 3 0} (u,\theta)} \vert^2$ for shell $s = 0.75$, $\alpha = 0.5$.}
		\label{fig:Dens7}
	\end{figure}

\begin{figure}
		\centering
		\includegraphics[width=0.6\columnwidth]{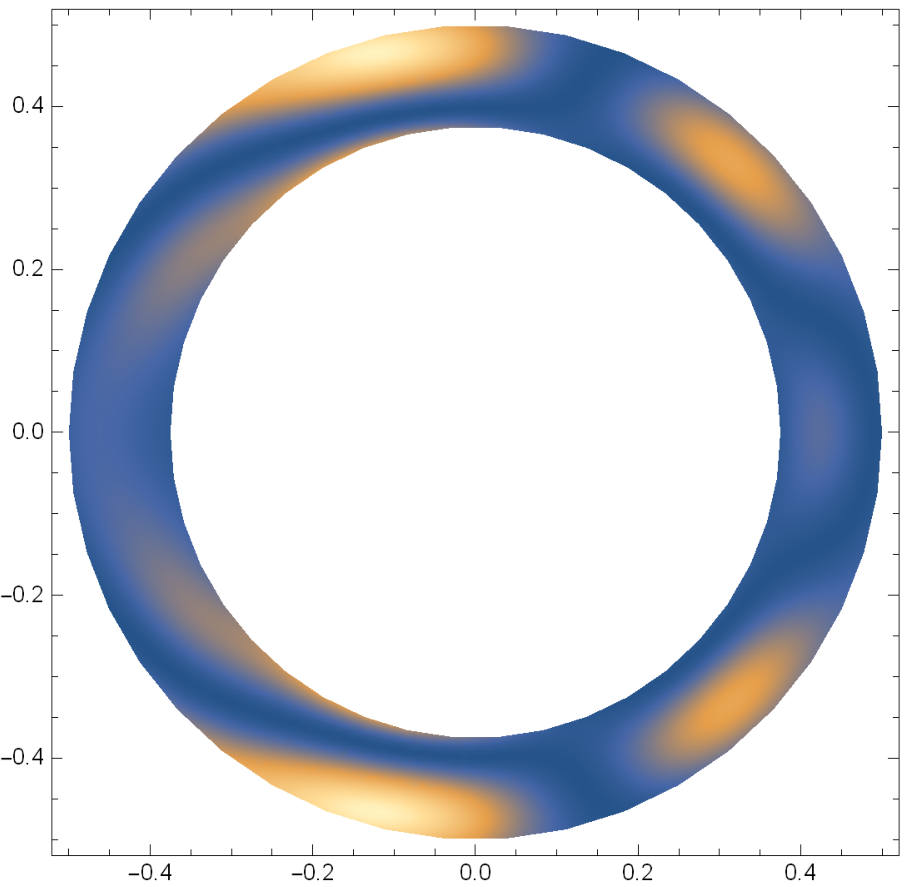}
		\caption{$\vert \psi{^{+}_{3 3 (10)} (u,\theta)} \vert^2$ for shell $s = 0.75$, $\alpha = 0.5$.}
		\label{fig:Dens8}
	\end{figure}

	Figures \,\ref{fig:Dens1}-\ref{fig:Dens4} show  volume and shell density plots for the  $n\nu m_{\phi} =010$ and $n\nu m_{\phi} = 01(10)$  states indicated in Tables 3 and 5 (in the interest of conciseness, only positive parity results are presented). Figures  \,\ref{fig:Dens5}-\ref{fig:Dens8}  present the same for $n\nu =33$ states.  A cross-sectional plot in general will not exhibit symmetry about $\theta = \pi / 2$; the larger radius of curvature at the inner rim leads to a left-right asymmetry for $\vert \psi{^{+}_{n \nu m_\phi}} \vert^2$ .  The results indicate increasing $m_\phi$ shifts densities rightward towards $\theta = 0$ mimicking a centrifugal force whose effect is more pronounced in shells than for volumes.

	The results above indicate that calculation of thermodynamic quantities requires a more comprehensive determination of the nano-scale torus spectrum than the method performed in \cite{WE}.  Additionally, the extent of mode coupling on the spectrum in that work was underestimated. Here, a sufficient number of eigenvalues, numbering in the several hundreds, are computed to insure convergence of the partition function
	
	\begin{equation} \label{eq:partition}
		Z(\tau)=\sum_N \exp[-{E_N}/{\tau}]
	\end{equation}
	
	\noindent
	up to 100 K. Results for only one shell with  $s = 0.75$ are given here, leaving a more detailed study of the interplay between aspect ratio and shell thickness for more extensive future analysis.  Further, in anticipation of applying the methods developed here to the case of the toroidal trap described in \cite{gold}, the dimensions of that trap $R = 875 \  {\rm\AA}, a = 250 \  {\rm\AA}$ are adopted in what follows.   
	
\begin{figure}
		\centering
		\includegraphics[width=0.75\columnwidth]{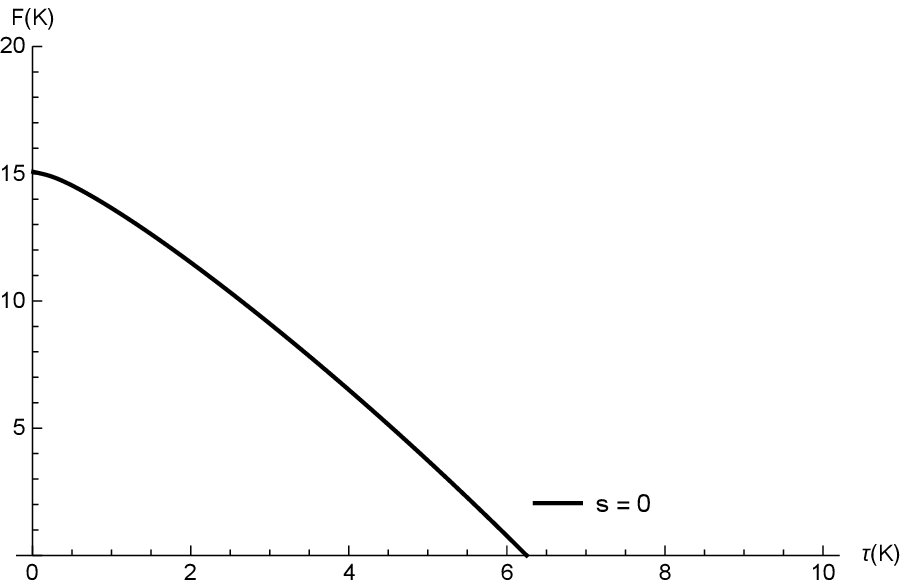}
		\caption{$F(\tau)$ as a function of temperature in Kelvin for $\alpha = 0.5$, $s = 0$.}
		\label{fig:F1}
	\end{figure}

\begin{figure}
		\centering
		\includegraphics[width=0.75\columnwidth]{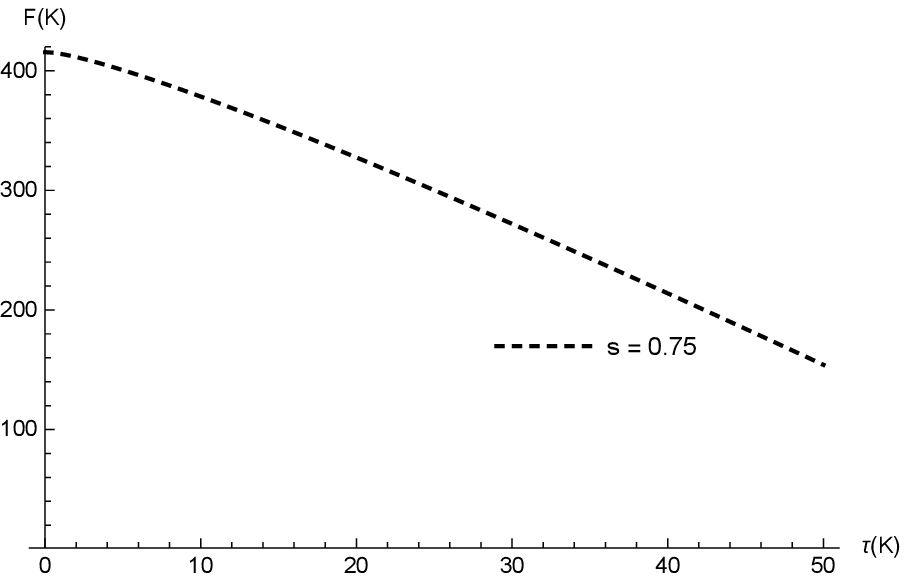}
		\caption{$F(\tau)$ as a function of temperature in Kelvin for $\alpha = 0.5$, $s = 0.75$.}
		\label{fig:F2}
	\end{figure}

	The effect of confinement to a shell is apparent in the results for thermodynamic variables. In Figure  \,\ref{fig:CV1} 
	the specific heat determined from 
	\begin{equation} \label{eq:specificheat}
		C_V(\tau)=-\frac{\partial U}{\partial \tau}
	\end{equation}
	indicates $C_V^{sh} > 	C_V^{vol}$ for $\tau < {\rm 5 \  K}$; this is a reflection of the energy scale of the shell being much larger than of the volume, as activation of the ground state for small but increasing $\tau$ yields a pronounced effect.  The plot shows that both  geometries converge to their assumed limits at approximately $30 {\rm \  K}$ with crossover at $5 \ {\rm  K}$. In Figures  \,\ref{fig:F1} and  \,\ref{fig:F2}, free energies  $F(\tau) = -\tau {\rm Ln}[Z(\tau)]$ are shown to contrast the scaling that arises from confinement.  The degeneracy pressure will also depend on the volume but is not immediately accessible from
	\begin{equation} \label{eq:pressure}
		p(\tau)=-\tau \frac{\partial F(\tau)}{\partial V}
	\end{equation}
	To arrive at a value for $p(\tau)$, first express the volume derivative as
	\begin{equation} \label{eq:chainruleV}
		\frac{\partial}{\partial V}= \frac{1}{2 \pi^2 R^3 \alpha } \frac{\partial }{\partial \alpha}
	\end{equation}
	\noindent
	for the full torus and
	\begin{equation} \label{eq:chainruleS}
		\frac{\partial}{\partial V}= \frac{1}{2 \pi^2 R^3 (1-s)^2 \alpha } \frac{\partial }{\partial \alpha}
	\end{equation}
	\noindent
	for the shell. With $R$ held fixed, five $a$ values corresponding to a $\Delta \alpha = 0.05$ were taken (in Angstroms,  $a =(162.49, 206.24, 250.00, 293.75, 337.49$) for $\tau = 1-5 {\rm \ K}$. A five point derivative formula and interpolating polynomial generated by Mathematica as a consistency check  were used to evaluate the derivative at $250 {\rm \ \AA}$ resulting in  pressure in the interval $0.01 {\rm \ K} \leq \tau \leq 5{\rm \ K}$,
	\begin{equation} \label{eq:pressureV}
		p_v(\tau)= 5.57(10^{-8}) {\rm \ K} / {\rm \AA^3}, 
	\end{equation}
	\begin{equation} \label{eq:pressureS}
		p_s(\tau)=1.27(10^{-7}) {\rm \ K} {\rm / \AA^3}
	\end{equation}
	
\begin{figure}
		\centering
		\includegraphics[width=0.75\columnwidth]{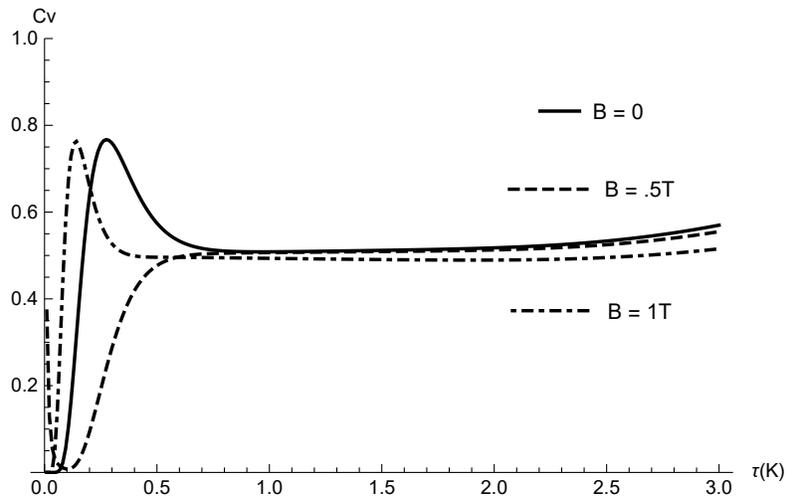}
		\caption{$C_V(\tau)$ as a function of temperature in Kelvin for $\alpha = 0.2875$, $s = 0$ for $B = 0, 0.5$ and $1.0 \ \rm T$.}
		\label{fig:S0CVs}
	\end{figure}

\begin{figure}
		\centering
		\includegraphics[width=0.75\columnwidth]{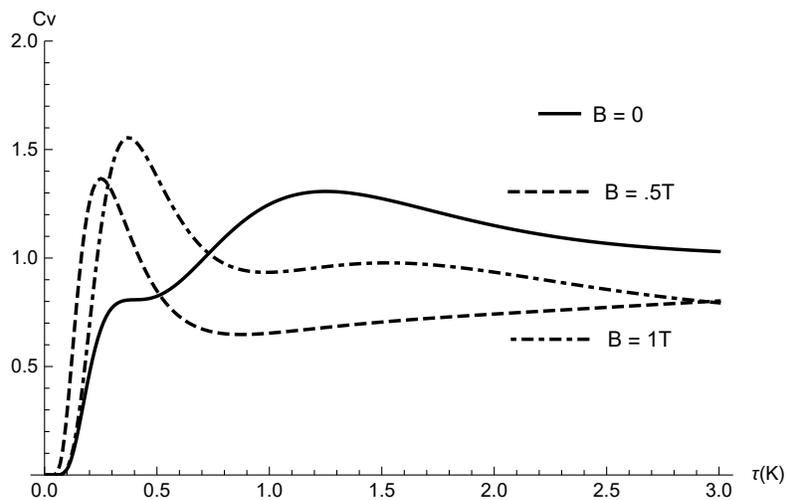}
		\caption{Heat capacity as a function of temperature in Kelvin for $\alpha =0.2875$,  $s = 0.75$ for for $B = 0, 0.5$ and $1.0 \ \rm T$. }
		\label{fig:S75CVs}
	\end{figure}

	Finally, the effect of a magnetic field on $C_V$ was not intended to be a primary focus of this work, but it bears noting that results are easily obtainable.  Figures \,\ref{fig:S0CVs} and \,\ref{fig:S75CVs}  demonstrate the sensitivity of the specific heat to moderate fields of the form $\bf B(r)$ =  $B_0\hat{{k}}$ at low temperatures.
	
	\section{Conclusions}
	The goal of this work in its preliminary stages was focused on the thermodynamics of electron gases in toroidal shells. It became apparent that more basis states and eigenvalues would be necessary than the number employed in \cite{WE} for a volume. The first half of this paper provided the methods by which those eigenvalues and wave functions were computed and indicates the spectrum strongly depends on $m_\phi$ in a way not captured by a simple approximation, thus requiring a reanalysis of earlier results. Once the wave functions were acquired, density plots for representative states provided visual insight into an effective centrifugal force causing shifts from the inner rim of the torus towards its outer rim for moderate values of $m_\phi$.
	
	With the eigenvalues acquired in sufficient number and accuracy, the latter half of this paper focused on  thermodynamic state functions with convergence questions within the temperature ranges considered here moot.  $F(\tau)$ and  $p(\tau)$ evidenced unsurprising behavior, however, the $C_V(\tau)$ plots are indicative of how confinement affects the magnitude and spacing of the spectrum and consequently, the specific heat.  
	
\begin{figure}
		\centering
		\includegraphics[width=0.75\columnwidth]{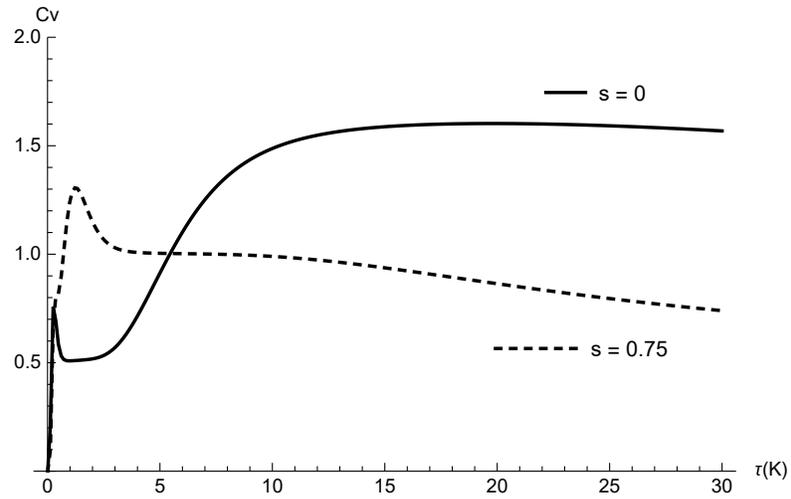}
		\caption{$C_V(\tau)$ as a function of temperature in Kelvin for $\alpha = 0.5$ with $s$ as indicated.}
		\label{fig:CV1}
	\end{figure}

	There are three length scales associated with a toroidal shell after re-scaling by $R$; $1, \alpha$, and $s\alpha$. For both volume and shell, the azimuthal energy scales as 1; however, the volume radial/poloidal energies scale as $1 / \alpha^2$ while for thin shells the radial energy scales as $1 / \alpha^2 (1 -s)^2$, and the nearly decoupled poloidal energy as  $1 / \alpha^2$.  Upon examining the eigenvalues it becomes clear for the volume case fewer $m_\phi$ are needed to match and surpass contributions made by the Bessel (radial/poloidal) $x_{n\nu}$ at  a given temperature.  For shells, however, fewer radial and poloidal contributing terms are required, but many more $m_\phi$  before there are roughly equivalent contributions to the particular energy of a state; this appears to be the origin of the $C_V$ behavior shown in Figure \,\ref{fig:CV1}.
	
	There are several directions for further investigation suggested by this work. As stated previously, conciseness motivated inclusion of only one shell thickness, chosen to approximate values of recently fabricated thin films \cite{film1}, but a  systematic study of aspect ratio and its interplay with shell thickness would give a measure of which variable is dominant in fixing the properties of a structure for given $R, a$ and $s$. Additionally, while proof of principle results were presented here for the influence of a magnetic field,  interesting results may emerge again depending on the values of $R, a$, $s$ with a more thorough treatment, particularly if vacuum in the region $0 \leq u \leq  s \alpha$ is replaced with a magnetic or conducting material. Finally, a description of gold toroidal traps by the methods presented here was thought to be prohibitive given the magnitude of eigenvalues required to reach the Fermi surface. However, results related to this work suggest that while a simple approximation method will not accurately represent the spectra, there may be sufficient predictability  within it to render another type of approximation feasible,  making   calculating persistent current and other low-temperature phenomena for metals with Fermi levels $\sim 5 \ e \rm V$ possible.

	\begin{table}
		\centering
		\resizebox{\columnwidth}{!}{
			\begin{tabular}[t]{c c c c c c c}
				\hline
				$s$ &\multicolumn{2}{c}{$.25$}&\multicolumn{2}{c}{$.5$}&\multicolumn{2}{c}{$.75$}\\
				$\eta_{n\nu}$ &Comp.&\multicolumn{1}{c}{Approx.} &Comp.&\multicolumn{1}{c}{Approx.}&Comp.&Approx.\\
				\hline
				$\eta_{01}$& 4.098& 4.189 & 6.246 & 6.283 & 12.553& 12.566\\
				$\eta_{12}$& 8.536& 8.378 &12.625 & 12.666 & 25.153& 25.132\\
				$\eta_{23}$& 13.121& 12.566 &19.045 & 18.850& 37.765& 37.699\\
				$\eta_{34}$& 17.747& 16.755 &25.477 & 25.133 &50.381& 50.265\\
				$\eta_{45}$& 22.389& 20.944 &31.912 & 31.416& 62.999& 62.831\\
				\hline
			\end{tabular}
		}
		\caption{Extended Bessel roots computed from Eqs. (11) and (12) compared to an approximate value for a hard wall of thickness 1-s. }
		\label{tab1}
	\end{table}
	
	\begin{table}
		\centering
		\resizebox{0.525\columnwidth}{!}{
			\begin{tabular}[t]{c c c c }
				\hline
				$s$ &\multicolumn{1}{c}{$.25$}&\multicolumn{1}{c}{$.5$}&\multicolumn{1}{c}{$.75$}\\
				$\psi_{n\nu}$ &\multicolumn{1}{c}{$B_{n\nu}$}&\multicolumn{1}{c}{$B_{n\nu}$}&\multicolumn{1}{c}{$B_{n\nu}$}\\
				\hline
				$\psi_{01}$& -7.045& 0.892 & 0.955\\
				$\psi_{12}$& 16.991& -0.838 &-0.933 \\
				$\psi_{23}$& 11.395& 1.871 &1.263 \\
				$\psi_{34}$& 14.130& -0.276 &-0.659\\
				$\psi_{45}$& 28.219& 23.6908 &1.859\\
				\hline
			\end{tabular}
		}
		\caption {Irregular Bessel function coefficient $B_{n\nu}$ indicated in Eq. (10) for $s = 0.25, 0.5$ and $0.75$ with $A_{n\nu}=1$}.
		\label{tab2}
	\end{table}
	
	\begin{table}
		\centering
		\resizebox{.85\columnwidth}{!}{
			\begin{tabular}[t]{l l}
				\hline
				$m_\phi$&$|\hspace{3em} \psi{^{+}_{0 1 m_\phi}} (u,\theta)$,\hspace{2em} $\alpha = 0.5,\hspace{2em} s = 0$\\ \hline
				$0$& $|-1.00 \chi{^{+}_{0 1 0}} (u,\theta)$\\
				
				$5$& $|\ \ \ \ 0.974\chi{^{+}_{0 1 5}} (u,\theta) + 0.222\chi{^{+}_{2 2 5}} (u,\theta)$\\
				
				$10$& $|-0.841\chi{^{+}_{0 1 5}} (u,\theta) - 0.527\chi{^{+}_{2 2 5}} (u,\theta)$\\
				\hline
			\end{tabular}
		}
		\caption{Full torus ground state from Eq.\,(\ref{eq:dimensionless}) for increasing $m_\phi$ neglecting terms of order $\sim  10^{-2}$.}
		\label{tab3}
	\end{table}
	
	\begin{table}
		\centering
		\resizebox{.9\columnwidth}{!}{
			\begin{tabular}[t]{l l}
				\hline
				$m_\phi$&$|\hspace{3em} \psi{^{+}_{3 3 m_\phi}} (u,\theta)$,\hspace{2em} $\alpha = 0.5,\hspace{2em} s = 0$\\ \hline
				$0$& $| \ \ \ \ 1.00 \chi{^{+}_{3 3 0}} (u,\theta)$\\
				
				$5$& $|\ \ \ \ 0.994\chi{^{+}_{3 3 5}} (u,\theta) - 0.256\chi{^{+}_{1 4 5}} (u,\theta)
				+ 0.055\chi{^{+}_{4 3 5}} (u,\theta)$\\
				
				$10$& $|-0.917\chi{^{+}_{3 3 5}} (u,\theta) - 0.527\chi{^{+}_{1 4 5}} (u,\theta) -0.209\chi{^{+}_{4 3 5}} (u,\theta)$\\
				\hline
			\end{tabular}
		}
		\caption{The (33) state evolution as $m_\phi$ is increased. }
		\label{tab4}
	\end{table}

	\begin{table}
	\centering
	\resizebox{.85\columnwidth}{!}{
		\begin{tabular}[t]{l l}
			\hline
			$m_\phi$&$|\hspace{3em} \psi{^{+}_{0 1 m_\phi}} (u,\theta)$,\hspace{2em} $\alpha = 0.5,\hspace{2em} s = 0.75$\\ \hline
			$0$& $|-1.00 \chi{^{+}_{0 1 0}} (u,\theta)$\\

			$5$& $|\ \ \ \ 0.743\chi{^{+}_{0 1 5}} (u,\theta) + 0.654\chi{^{+}_{1 1 5}} (u,\theta)- 0.141\chi{^{+}_{2 1 5}} (u,\theta)$\\
			
			$10$& $|\ \ \ \ 0.639\chi{^{+}_{0 1 5}} (u,\theta)+ 0.697\chi{^{+}_{1 1 5}} (u,\theta)-0.315\chi{^{+}_{2 1 5}} (u,\theta)$\\
			\hline
	\end{tabular}}
	
	\caption{Shell s = 0.75 ground state from Eq.\,(\ref{eq:dimensionless}) for increasing $m_\phi$ neglecting terms of order $\sim  10^{-2}$.}
	\label{tab5}
\end{table}

\begin{table}
	\centering
	\resizebox{1.1\columnwidth}{!}{
		\begin{tabular}[t]{l l}
			\hline
			$m_\phi$&$|\hspace{3em} \psi{^{+}_{3 3 m_\phi}} (u,\theta)$,\hspace{2em} $\alpha = 0.5,\hspace{2em} s = 0.75$\\ \hline
			$0$& $|-1.00 \chi{^{+}_{3 3 0}} (u,\theta)$\\
			
			$5$& $| -0.232\chi{^{+}_{0 3 5}} (u,\theta) + 0.366\chi{^{+}_{1 3 5}} (u,\theta)
			+0.481\chi{^{+}_{2 3 5}} (u,\theta) - 0.641\chi{^{+}_{4 3 5}} (u,\theta)+0.411\chi{^{+}_{4 3 5}} (u,\theta)
			$\\
			
			$10$& $|\ \ \  0.319\chi{^{+}_{0 3 (10)}} (u,\theta) - 0.463\chi{^{+}_{1 3 (10)}} (u,\theta)
			-0.489\chi{^{+}_{2 3 (10)}} (u,\theta) + 0.502\chi{^{+}_{4 3 (10)}} (u,\theta)+0.438\chi{^{+}_{4 3 (10)}} (u,\theta)$\\
			\hline
		\end{tabular}
	}
	\caption{Shell s = 0.75 ground state from Eq.\,(\ref{eq:dimensionless}) for increasing $m_\phi$ neglecting terms of order $\sim  10^{-2}$.}
	\label{tab6}
\end{table}

	\section{Appendix}
	\renewcommand{\theequation}{A.\arabic{equation}}
	\setcounter{equation}{0}  
	Unit vectors cited in section 2 are
	
	\begin{align}
		\hat{\textbf n} &= \cos\theta \, \hat{\boldsymbol \rho} + \sin \theta \,\hat{\textbf{k}}\\
		\hat{\boldsymbol \theta} &= -\sin\theta \,\hat{\boldsymbol \rho} + \cos \theta \,\hat{\textbf{k}}\\
		\hat{\boldsymbol \phi} &= -\sin \phi \,\hat{\textbf{i}} + \cos \phi \,\hat{\textbf{j}}\,
	\end{align}

	The $\Lambda_{\bar{n} n}^{\pm}(u)$ are generated from 
	\begin{align}
		\Lambda_{\bar{n} n}^{+}(u) &= \int_{0}^{2\pi} d\theta \frac { \cos \bar{n}\theta \cos \theta  \cos n\theta}
		{(1 + u\cos\theta)^2}\\
		\Lambda_{\bar{n} n}^{-}(u) &= \int_{0}^{2\pi} d\theta \frac { \sin \bar{n}\theta \sin \theta  \sin n\theta}
		{(1 + u\cos\theta)^2}
	\end{align}

	Let $w(u)=\sqrt{1-u^2}$ and define
	\begin{align}
		I_1 &=\frac {1} {[w(u)]^{3}} \big [- \frac {1} {u}(1 - w(u)) \big ]^{\bar{n}+n} \big [( {\bar{n}+n})w(u)+1 \big ]\\
		I_2 &=   \frac {1} {[w(u)]^{3}} \big [-\frac {1} {u}( 1 - w(u)) \big ]^{|{\bar{n}-n}|} \big [( |{\bar{n}-n})|w(u)+1 \big ]
	\end{align}
\noindent 
Eqs. (A.6) and  (A.7) are corrected forms of the corresponding expressions in \cite{WE}.
	Then
	\begin{equation}
		\Lambda_{\bar{n} n}^{\pm}(u) = \pi (I_2 \pm I_1)
	\end{equation}
	
	\section{Acknowledgements}
	M.E. would like to thank Lewis Johnson for providing software support for this work.
	\bibliographystyle{elsarticle-num}
	\biboptions{sort&compress}
	\bibliography{referencebase3}
\end{document}